# Electrically-Pumped Wavelength-Tunable GaAs Quantum Dots Interfaced with Rubidium Atoms

Huiying Huang,[1, 2] Rinaldo Trotta,*[1] Yongheng Huo, *[1,2] , Thomas Lettner,[1] Johannes S. Wildmann,[1] Javier Martín-Sánchez,[1] Daniel Huber,[1] Marcus Reindl,[1] Jiaxiang Zhang,[2] Eugenio Zallo,[2,3] Oliver G. Schmidt,[2] and Armando Rastelli[1]

[1]Institute of Semiconductor and Solid State Physics, Johannes Kepler University, Linz, Altenbergerstraße 69, 4040, Austria

[2]Institute for Integrative Nanosciences, IFW Dresden, Helmholtzstraße 20, 01069 Germany

[3]Paul-Drude-Institut für Festkörperelektronik Hausvogteiplatz 5-7, 10117 Berlin, Germany

**ABSTRACT:** We demonstrate the first wavelength-tunable electrically-pumped source of non-classical light that can emit photons with wavelength in resonance with the $D_2$ transitions of $^{87}$Rb atoms. The device 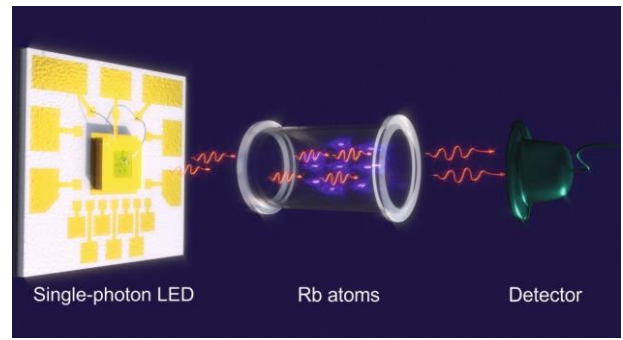 is fabricated by integrating a novel GaAs single-quantum-dot light-emitting-diode (LED) onto a piezoelectric actuator. By feeding the emitted photons into a 75-mm-long cell containing warm $^{87}$Rb atom vapor, we observe slow-light with a temporal delay of up to 3.4 ns. In view of the possibility of using $^{87}$Rb atomic vapors as quantum memories, this work makes an important step towards the realization of hybrid-quantum systems for future quantum networks.

Single photon sources (SPSs) play a key role in quantum communication and quantum computation.[1, 2] For practical applications, it would be highly desirable to have electrically–pumped SPSs that are compatible with other key elements of future quantum networks, such as quantum memories and single photon detectors.[3, 4]

Single semiconductor quantum dots (QDs), or "artificial atoms", are arguably among the most promising SPSs because of their capability of emitting single photons with high efficiency,[5] at high speed,[6] and with large degree of indistinguishability.[7, 8] Furthermore, semiconductor QDs can be embedded into the active region of light-emitting-diodes[9, 10, 11, 12, 13] and their optical properties can be finely adjusted using external perturbations.[14,15,16,17]

The combination of artificial and natural atoms, a research avenue pioneered by Akopian et al.,[18] may allow storing the state of single photons emitted by a QD in atomic vapours or Bose-Einstein-Condensates.[19] Such atomic systems offer in fact particularly long storage times,[3] and may become crucial ingredients for building up a quantum repeater.[4, 20] One of the key requirements to build up this hybrid artificial-natural atomic system is the availability of QDs with emission precisely tuned to atomic transitions. In the first work[18] – focused on GaAs/AlGaAs QDs interfaced with $^{87}$Rb atomic vapours – energy tuning and QD excitation were achieved through magnetic fields and an external laser source, respectively. This makes the overall single-photon source extremely bulky and impractical for the envisioned hybrid quantum networks.

In this work, we report on the very first realization of a wavelength-tunable GaAs QD based quantum-light-emitting diode (Q-LED), an all electrically-driven source of single photons (see Fig. 1) in which both the excitation and wavelength-tunability are provided on-chip while preserving the compatibility with the $^{87}$Rb $D_2$ line transitions. To demonstrate the suitability of the device for future applications based on hybrid QD-atomic systems, we send the emitted light into a $^{87}$Rb-vapour cell and observe slow single photons when the QD transition energy is fine-tuned through the hyperfine-split D$_2$ lines of Rb.

The wavelength-tunable Q-LED is obtained by integrating nano-membrane p-i-n diodes on a [Pb(Mg$_{1/3}$Nb$_{2/3}$)O$_3$]$_{0.72}$-[PbTiO$_3$]$_{0.28}$ (PMN-PT) piezoelectric actuator, as sketched in Fig. 1. It is important to note that, while single-QD-LEDs have been reported for other material systems, such as InGaAs/GaAs,[6,9,10,11,13] and InP/InGaP,[12] reports on electrically-controlled GaAs QDs are limited to n-i-Schottky structures,[21] which are unsuitable for electrical injection of both electrons and holes. The Q-LED nano-membrane used here was grown by molecular beam epitaxy (MBE) and consists of a QD layer in the middle of a 124 nm Al$_{0.4}$Ga$_{0.6}$As intrinsic layer, which is sandwiched between a 160 nm n-doped Al$_{0.4}$Ga$_{0.6}$As layer and an 85 nm p-doped Al$_{0.4}$Ga$_{0.6}$As layer, see Fig. 1. The QDs were obtained by an optimized local-droplet-etching approach consisting in Al-droplet etching[22] of an Al$_{0.4}$Ga$_{0.6}$As layer followed by nanohole filling with GaAs.[23,24] By adjusting the GaAs filling amount, the QD emission wavelength can be tuned to be around 780 nm,[25] close to the $^{87}$Rb D$_2$ transitions. The heterostructure was capped by 10 nm highly doped GaAs on both sides to prevent oxidation and to enable ohmic contacts. QDs grown with this technique can have a density lower than $10^8$ cm$^{-2}$, making it easy to address single dots in micro-electroluminescence (μ-EL) measurements. Recent studies[26,27] have also shown that such QDs can emit highly indistinguishable and entangled photons, making them appealing for envisioned quantum-optical experiments and applications.

The active structure was grown on a sacrificial layer, which allows free-standing membranes to be fabricated and integrated onto a 300 μm thick PMN-PT piezoelectric actuator via flip-chip and gold thermo-compression bonding.[13] The device was mounted onto an AlN chip carrier providing electrical contacts both to the diode nanomembranes and the piezoelectric actuator. The current is injected into the Q-LED by applying a bias voltage V$_d$ above the turn-on voltage of the diode, while the strain-state of the membrane is varied by applying a voltage $V_p$ across the PMN-PT piezoelectric actuator, as illustrated in Fig 1. The two circuits share a common ground and allow independent control of LED current - and thus emission intensity - and strain, which in turn controls the emission wavelength due to strain-induced effects.[6, 13, 16, 17, 28] The thickness of the layers and the position of the QD is chosen to obtain a simple Au-semiconductor-air planar cavity, which enhances the collection efficiency of the QD emission.[6, 13]

In order to measure the EL spectra of the single Q-LED, the device is mounted on the cold-finger of a helium-flow cryostat and cooled to temperatures of ~10 K. Fig. 2a displays a typical I-V trace of the Q-LED, showing no observable current below the turn-on voltage and a rapid increase beyond it, as in standard LEDs. (Note that the sign of the voltage needed to obtain EL is negative, since the top contact of the diode is n-type. In Fig. 2 and in the following we omit the minus sign on both the voltage and the current). To collect EL from single QDs, a 50× microscope objective with a numerical aperture of 0.42 is used. The signal is then guided into a spectrometer equipped with an 1800 lines/mm grating and a liquid-nitrogen-cooled Si-CCD. The spectrometer has a resolution of about 40 μeV at the wavelength of 780 nm. In Fig. 2b, we show the bias-voltage dependent μ-EL map from a single QD. When the bias voltage $V_d$ exceeds the turn-on voltage of the diode (about 2.2~2.3 V), photons of different wavelengths are emitted due to the recombination processes of the injected electrons and holes. The linewidth of the dominant neutral exciton state (X) is below the resolution limit of the setup. It is important to note that in the explored range of $V_d$ only the line intensity changes, while no appreciable spectral shift nor line-broadening is observed (see Fig. 2b).

To interface with Rubidium atoms, we insert a Rb gas cell in the optical path, between the objective and the spectrometer. The cell, made of quartz and with a length of 75 mm, contains mostly $^{87}$Rb atoms, with some residual of $^{85}$Rb atoms. A heating setup is used to change the vapor density of the cell. A temperature calibration of the heating setup can be found in the supplemental materials. As in ref.18, we use the double absorption resonance scheme to demonstrate slow single photons.[3, 18, 29] Specifically, we tune the energy of the photons in the spectral range of the D$_2$ transitions (5S$_{1/2}$ to 5P$_{3/2}$) of $^{87}$Rb (~780 nm), whichconsist of 6 lines in total.[30] However, limited by the linewidth of the QD emission, the Doppler broadening of the atomic lines, and the spectrometer resolution, we can resolve only the two subgroups separated by the hyperfine splitting of the ground state 5 S$_{1/2}$ of about 28 μeV.

The emission wavelength of the neutral exciton X (the dominant exciton's transition in Fig. 2b) is scanned through the D$_2$ transitions by gradually varying the electric field applied to the PMN-PT piezoelectric actuator, as shown in the color-coded μ-EL spectra of Fig. 2c. Two transmission dips are clearly observed at F$_p$= 16.05 and 16.40 kV/cm due to the absorption at the D$_2$ transitions. To better observe this effect, we extract from each spectrum the peak intensity (without background) of the X line through a Lorentzian fit. The result is shown in Fig. 2d, where the two intensity dips corresponding to the hyperfine splitting of the ground state are clearly identified. Please note that, even at these two dips, the transmission intensity is not zero, mostly due to the finite linewidth of the X emission. The energy difference between the two dips is found to be ~ 27 μeV, thus matching the expectations.[30]

When the emission energy of the photons emitted by the Q-LED is set to the range of the absorption doublet, we expect a temporal delay for the transmitted photons due to the strongly dispersive behavior of the vapor medium,[29, 31, 32] which in turn leads to a decrease of the light group velocity as discussed in the supplemental materials.

To observe such a delay and prove the non-classical nature of the light emitted by our GaAs QD-LED, we measured the second-order autocorrelation function $g^{(2)}(\tau)$ of photons emitted by the X recombination of another - brighter - QD using a Hanbury-Brown-Twiss (HBT) setup. This X transition (see Fig. 3a) consists of two orthogonally polarized lines separated by a fine-structure-splitting of 145 µeV. Only the brightest polarization component, with a linewidth of 58 µeV is selected for the HBT measurements. As sketched in Fig. 3b, the µ-EL beam is separated by a 50/50 beam splitter and sent to two avalanche photodiodes (APDs) via the two arms of the HBT setup. The time-resolution of the setup is about 900 ps, as determined from an autocorrelation measurement of femtosecond laser pulses. In one of the paths, we placed the $^{87}$Rb cell. Finally, the two APDs outputs are connected to correlation electronics to record coincidence counts. Results (with no background correction) are shown in Fig. 3c. The black lines in Fig. 3c show the measured $g^{(2)}(\tau)$ when the QD-photons are tuned off resonance with respect the $^{87}$Rb $D_2$ transitions. The second-order correlation function at zero time delay $g^{(2)}(0)$ reaches a value of 0.20. The red and blue lines show the $g^{(2)}$ results when the emission wavelength is turned on resonance with the $D_2$ transitions at Rb cell temperatures $T_{Rb}$ of 96 (red) and 100 °C (blue). Compared with the off-resonance autocorrelation result, we observe two effects with increasing $T_{Rb}$: (1) the position of the minimum of the $g^{(2)}(\tau)$, $\tau_c$, shifts from 0 to 1.7 ns (at $T_{Rb}$ = 96 °C) and 3.4 ns (at $T_{Rb}$ = 100 °C); (2) The values of $g^{(2)}(\tau_c)$ increase from 0.2 to 0.42 (at $T_{Rb}$ = 96 °C) and 0.51 (at $T_{Rb}$ = 100 °C), as indicated in Fig. 3c.

The dips in all three measurements clearly indicate that our device is capable of emitting non-classical states of light. The temporal shifts of the minima of $g^{(2)}(\tau)$ in resonant conditions can be easily explained by the reduction of group velocity of photons tuned between the absorption lines of Rb[33]. The increase of $g^{(2)}(\tau_c)$ is attributed to the strong optical dispersion of the atomic medium combined with the finite linewidth of the QD emission (58 µeV).[32, 34] More precisely, since the group velocity is wavelength-dependent, QD photons of slightly different energy traverse the vapour with different group velocities and thus escape the gas cell at slightly different times. In turn, the non-ideal linewidth of the X emission of the selected QD is attributed to spectral jitter produced by fluctuating electric fields at the QD position. This spectral jitter, combined with the spectral dependence of the group velocity, leads to a time jitter in the arrival times of photons on the APDs. This time jitter (which increases with increasing cell temperature $T_{Rb}$) adds up to the instrumental jitter of the APDs leading to a broadening and smearing of the $g^{(2)}(\tau)$ dip. We model the dip position shifts by introducing a temporal shift $\tau_c$ in the formula for the ideal second order correlation function of a two level system:

$$g^{(2)}(\tau) = 1 - exp\left(\frac{|\tau-\tau_c|}{t}\right)$$

where *t* accounts for exciton relaxation and decay rates. In the equation above we have assumed a negligibly small electron-hole pair generation rate.[35]

To take into account the effect of time-jitter we perform a convolution with a Gaussian function

$$G(\tau) = \left[w\sqrt{\frac{\pi}{4\ln(2)}}\right]^{-1} exp\left(-\frac{4\ln(2)\tau^2}{w^2}\right)$$

whose FWHM *w* takes into account both the finite time resolution of the single photon detectors ($\Delta t$) and the time jitter introduced by the dispersive atomic medium ($\Delta t_c$). By assuming Gaussian statistics and considering that these two jitter mechanisms are uncorrelated, we have $w^2 = \Delta t^2 + \Delta t_c^2$.

After convolution the equation becomes

$$F(\tau) = g^{(2)}(\tau) * G(\tau)$$

where * indicates the convolution operation.

To reproduce the experimental data in the absence of the gas cell (or for QD emission far detuned from the atomic transitions), only the decay rate $t$ is used as free parameter (since $\Delta t$ is fixed at the experimental value, $\Delta t_c = 0$ and $\tau_c = 0$). The smooth curves in Fig. 3c show, for each $g^{(2)}(\tau)$ measurement, the corresponding fitting results. The decay time is found to be $t = 2.0$ ns, an extremely long value that we attribute to slow relaxation dynamics of carriers from the high energy states (see ref. 26). In the presence of the Rb cell at 96 °C, a temperature-dependent temporal shift of $\tau_c(96\ °C)= 1.7$ ns, time jitter of $\Delta t_c(100°C)= 4.1$ ns are instead needed to reproduce the experimental data. At $T_{Rb}=100$ °C, the values are $\tau_c(100\ °C) = 3.4$ ns and $\Delta t_c(100\ °C)= 5.7$ ns. It is worth emphasizing that the increased value of the $g^{(2)}(\tau_c)$ cannot be accounted for, as expected, by any artificial change in $t$, but it is totally ascribable to the additional time jitter $\Delta t_c$ discussed above.

In conclusion, we have reported on the first quantum light emitting diode based on single GaAs/AlGaAs QDs and demonstrated its operation as an energy-tunable source of photons in the spectral range of the $^{87}$Rb D$_2$ transitions. Autocorrelation measurements under continuous excitation clearly demonstrate antibunched emission, which is typical for single-photon sources. By comparing autocorrelation measurements with emission on/off resonance with the $^{87}$Rb D$_2$ transitions we deduce that a double-resonance slow-light mechanism produces temporal delays of up to 3.4 ns. To increase the delay, quantum dot samples with even narrower emission lines would be required. Then the delay can be further extended by increasing the vapor density in the Rb cell. Further optimization of the LED design shall enable pulsed operation and increased extraction efficiency. We envision that this all-electrically controlled single photon LED, once integrated with chip-scale atom quantum memory such as hollow-core fiber cell,[36-38] may pave the way towards large scale small volume hybrid quantum systems that can be used in future quantum networks.

**FUNDING**

This work is financially supported by BMBF QuaHL-Rep (Contracts no. 01BQ1032 and 01BQ1034), the European Union Seventh Framework Program 209 (FP7/2007-2013) under Grant Agreement No. 601126 210 (HANAS).

**AUTHOR INFORMATION**

**Corresponding authors:**

*rinaldo.trotta@jku.at

*yongheng.huo@jku.at

**Notes**

The authors declare no competing financial interest

**ACKNOWLEDGMENT**

The authors thank B. Eichler, R. Engelhard, S. Harazim, F. Binder, A. Halliovic, U. Kainz, E. Vorhauer, and S. Bräuer for technical assistance and B. Höfer, V. Mahalingam, and F. Scholz for helpful discussions.

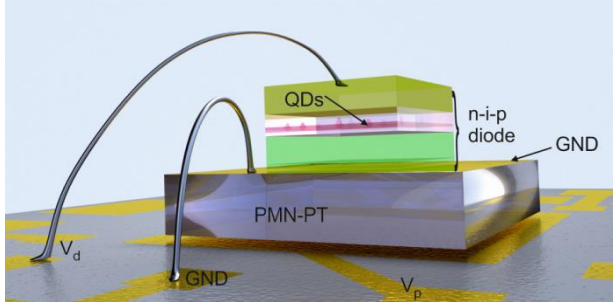

**Figure 1.** Sketch of the electrically-pumped wavelength-tunable Q-LED. The n-i-p diode contains GaAs QDs embedded in $Al_{0.4}Ga_{0.6}As$ barriers and is integrated on a PMN-PT piezoelectric actuator which provides variable strain fields to tune the photon emission energy. $V_d$ is the bias voltage applied to the diode, $V_p$ is the voltage applied to the actuator.

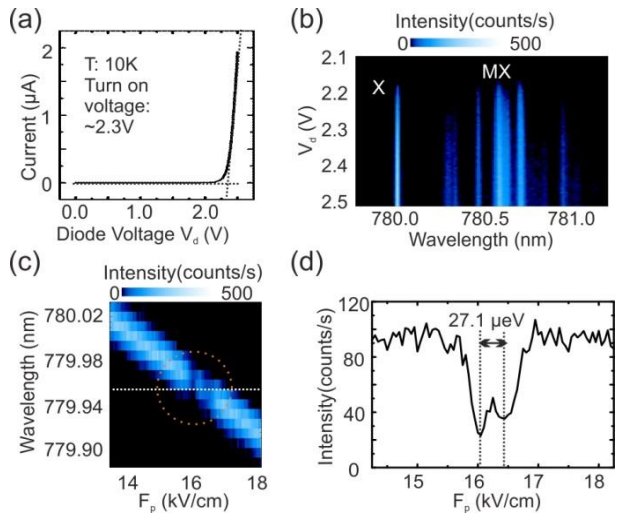

**Figure 2.** (a) Typical current-voltage (I-V) characteristics of the QD LED, with a diode turn-on voltage of ~2.3 V. (b) Evolution of the electroluminescence (EL) spectra of a single QD embedded in the tunable-LED with the magnitude of the applied voltage $V_d$. The intensity (in CCD counts per second) is color-coded. In the spectra, the brightest line stems from the neutral exciton (X) transition, which is well separated from the group of low-energy states, which are ascribed to charged and neutral multi-excitonic states. (c) Color-coded micro-electroluminescence (µ-EL) spectra of the exciton in (a), whose energy is scanned across the $D_2$ transitions of the $^{87}$Rb cloud (at a temperature $T_{Rb}$=70°C) by applying variable stress on the Q-LED membrane. The white dotted line indicates the middle wavelength of the $D_2$ transitions. The transmitted intensity drops at 16.05 kV/cm and 16.40 kV/cm due to the absorption by the atomic vapor (region indicated by dashed circle). (d) Transmission intensity as a function of the electric field applied to the piezoelectric actuator ($F_p$), obtained by extracting the peak intensity values from the corresponding spectra in (c), as described in the text. The two dips correspond to the two hyperfine lines of the ground state of the $D_2$ transitions.

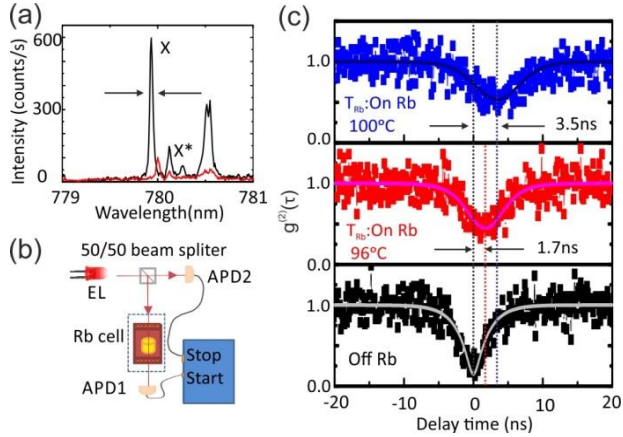

**Figure 3**. (a) EL spectra of a second QD, acquired with a double spectrometer equipped with two 1200 l/mm gratings (spectral resolution ~30μeV). The two spectra correspond to two orthogonal polarization directions. Only the bright line marked by arrows (with a linewidth of 58 μeV) is used in the following measurement. (b) HBT set-up used to extract the second-order correlation function, $g^{(2)}(\tau)$, of the EL emitted by a single QD embedded in the Q-LED. A 75 mm long $^{87}$Rb vapour cell is inserted in one of the arms of the setup. (c) Black: $g^{(2)}(\tau)$ measurements of the EL emission when the photon energy is tuned off resonance with respect

to the $^{87}$Rb D$_2$ transitions. Red and blue: $g^{(2)}(\tau)$ values when the photon energy is tuned on resonance with the $^{87}$Rb D$_2$ lines and the Rb cell temperature is 96 (red), and 100 °C (blue). The corresponding minimum values of $g^{(2)}(\tau)$, reached at a $T_{Rb}$-dependent delay time $\tau_c(T_{Rb})$ are 0.20, 0.42 and 0.51 for the data displayed in of the black, red and blue, respectively. The gradual increase of $\tau_c$ and of $g^{(2)}(\tau_c)$ with increasing $T_{Rb}$ comes from the dispersion of the atomic optical medium. This leads to slow light and, due to the finite linewidth of our photon source, to time-jitter and consequent broadening of the $g^{(2)}(\tau)$ curve.